\documentstyle[11pt,aaspp4,epsf]{article}

\begin{document}

\title{Image Deconvolution of the Radio Ring PKS~1830--211\altaffilmark{1}.}

\author{F.    Courbin}
\affil{Institut   d'Astrophysique   et     de
G\'eophysique, Universit\'e  de  Li\`ege,\\
 Avenue de  Cointe 5, B--4000 Li\`ege, Belgium\\ 
URA 173 CNRS-DAEC,  Observatoire de Paris, F--92195
Meudon Principal C\'edex, France}
\author{C.  Lidman}
\affil{European Southern Observatory, Casilla 19001, 
Santiago 19, Chile}
\author{B. L. Frye}
\affil{Astronomy Department, University of California,
    Berkeley, CA 94720, USA}
\author{P. Magain\altaffilmark{2}}
\affil{Institut   d'Astrophysique   et     de
G\'eophysique, Universit\'e  de  Li\`ege,\\
Avenue de  Cointe 5, B--4000 Li\`ege, Belgium}  
\author{T. J. Broadhurst}
\affil{Astronomy Department, University of California,
    Berkeley, CA 94720, USA}
\and
\author{M. A. Pahre\altaffilmark{3,4} and S. G. Djorgovski} 
\affil{Palomar Observatory, California Institute of
Technology, Mail Stop 105-24, Pasadena, CA 91125, USA}



\altaffiltext{1}{Based on observations obtained at the ESO La Silla
Observatory, Chile and at the W. M.  Keck Observatory, Hawaii, 
which is operated
jointly by the California Institute of Technology and the University of
California.}  \altaffiltext{2}{Ma\^{\i}tre de Recherches au FNRS (Belgium)}
\altaffiltext{3}{Present address:  Harvard-Smithsonian Center for Astrophysics,
    60 Garden Street, Mail Stop 20, Cambridge, MA  02140, USA}
\altaffiltext{4}{Hubble Fellow}




\begin{abstract}

New high  quality Keck and ESO  images of PKS~1830--211 are presented.
Applying  a powerful new  deconvolution algorithm to  these optical and
infrared  data,  both images  of the flat  spectrum  core of the radio
source  have been identified.  An  extended source is also detected in
the  optical  images, consistent  with the  expected  location of  the
lensing  galaxy.  The source counterparts  are very red at $I-K\sim7$,
suggesting strong  Galactic absorption  with additional absorption  by
the lensing galaxy at $z=0.885$,  and consistent with the detection of
high redshift molecules in the lens.

\end{abstract}




\section{Introduction}

   The bright radio   source PKS~1830--211 (Subrahmanyan et al.  1990;
hereafter S90, Jauncey  et al.  1991) has  attracted much attention as
the  most   detailed  example  of  a lensed   radio   ring.  Among the
classically-lensed QSOs, its short time delay of 44 days (van Ommen et
al.   1995) and  clean  lens geometry  make it  a  good candidate  for
measuring  H$_{0}$.   The lens,  a  gas rich   galaxy at  z=0.89,  was
discovered  in  the  millimeter via  molecular  absorption (Wiklind \&
Combes 1996), which is seen towards only one  of the two flat spectrum
hot spots (Wiklind \& Combes 1996, Frye et al. 1997).  The presence of
a  nearby M-star as  well as heavy extinction along  the line of sight
(b=-5.7 degrees) had until now hampered the identification of the lens
and the source.   In this paper  we describe how the MCS deconvolution
algorithm (Magain, Courbin,  \& Sohy   1998)  was used to detect   the
counterparts of   this bright radio   ring in  deep Keck  optical  and
infrared images.

\section{Observations - Reductions}

Near      IR  $J$   ($\lambda_c$=1.25    micron)   and    $K^{\prime}$
($\lambda_c$=2.15 micron) images were  obtained on the nights of  1997
April 14 and 15 with the IRAC2b camera on  the ESO/MPI 2.2m telescope,
which  uses a  NICMOS3  256$\times$256 HgCdTe array.   The good seeing
(0\farcs6-0\farcs7) and the fact that a good sampling is desirable for
deconvolution, led us to choose  the smallest pixel size available  on
this instrument, i.e., 0\farcs151, resulting  in a total field of view
of 38\arcsec\, on a side.  The data were  taken and reduced exactly in
the same way as in Courbin, Lidman \& Magain (1998).  Several standard
stars were  observed during the night.   The standard deviation in the
zero  points  was  less than   0.02  magnitudes    for both $J$    and
$K^{\prime}$.  The  IR magnitudes  reported in this  paper are  on the
$JHK$ system of Bessell \& Brett (1988).

Near-IR Keck~{\sc i} data were obtained  on the night  of 1994 April 5
with NIRC (Matthews \& Soifer, 1994).  Five 80  second $K$ images were
combined using the  sigma clipping algorithm available  in  IRAF.  The
pixel size is 0\farcs157, similar to that used with IRAC2b. The images
were obtained under marginally non-photometric conditions, with a zero
point uncertain by  about 0.1 magnitude.  Due  to the  crowding of the
field, and the low  number  of frames  available, sky subtraction  was
imperfect  but did  not   influence dramatically  the quality of   the
data analysis.

Six dithered $I$-band images were obtained during  the full moon night
of 1997 June 15 using the Keck~{\sc ii} telescope and LRIS (Oke et al.
1995).  The CCD detector was a Tektronix 2048$\times$2048 with a pixel
size of 0\farcs215.  The individual exposure  times were restricted to
3 minutes in order to avoid saturation of  the brightest stars in this
extremely  crowded field.  The background  was  very high.  The images
were bias subtracted  and flat fielded  in the  standard way.   All of
them have a seeing close to 0\farcs8.  No  standard star was observed,
but  a  flux calibration could be   performed relative to  an $I$-band
image taken at La Silla with the  0.9m telescope on  the night of 1997
April 15.

\section{Image deconvolution}

The MCS deconvolution code (Magain et al. 1997) was applied to all the
images.   Due  to the low  signal-to-noise  (S/N) and the numerous bad
pixels  in  single IR images,  these  were  medianed  in  groups of  9
dithered and sky subtracted frames.  The  resulting images have better
S/N and cosmetics.  Two infrared-bright nearby stars, although outside
the field of view,  produced scattered light in  the field, forcing us
to reject a fair fraction ($\sim 40 $ percent)  of our observations in
the $K^{\prime}$ band.  One of the  culprits was the known source IRAS
18306-2106.  Two stacks were obtained  in $J$ (total exposure time  of
1920 sec) and four in $K^{\prime}$ (total  exposure time of 1440 sec).
Only one such stack  was obtained for the  IR Keck images since we had
only 5 dithered frames to combine (total exposure time of 400 sec).

\subsection{Application of the MCS code to the present data}

The deconvolution process is the same as  in Courbin et al. (1998). We
chose a pixel scale in the deconvolved images  that is a factor of two
smaller than the pixel scale in the original  data, to insure that the
sampling   theorem  is satisfied  in   the  deconvolved images.  After
deconvolution,   the       resolution  is  fixed     to     2   pixels
Full-Width-Half-Maximum (FWHM) for  all data.  The corresponding final
pixel  scales and  resolutions on the  plane of  the  sky are given in
Fig. 1.

We   constructed  the infrared   Point-Spread-Function  (PSF) from the
relatively  isolated,    bright  star  labelled  H    in Djorgovski et
al. (1992; hereafter D92).  In the optical images, the stars mentioned
in D92 are all saturated.   Consequently, the PSF was constructed from
4 fainter stars about 30\arcsec\  away from the  radio ring.  Crowding
in the  optical   field made the   quality  of the PSF  rather   poor,
especially in  the outer wings of the  hexagonal Keck PSF, introducing
systematic residuals into the deconvolved image and residual maps (see
section 3.2).

In each   band,   all  the  frames  available    were {\it deconvolved
simultaneously}, i.e.  6 in $I$, 2 in $J$, 4 in $K'$  and 1 in $K$. In
other words, the output of the procedure is a unique deconvolved image
which is simultaneously compatible with all the images included in the
data set.  The  deconvolved image is  given as a  sum of point sources
with {\it known Gaussian  shape} and a  diffuse background smoothed on
the  length scale of   the final resolution  chosen  by the user.  The
photometry and  the astrometry of the point  sources are also obtained
as byproducts of the deconvolution and are provided in Tables 1 and 2.

\subsection{Quality check of the deconvolution}

Many deconvolution  algorithms generate  the so-called ``deconvolution
artefacts''  as well as noise enhancement.   Even if the MCS code does
not produce  artefacts, it still has to  accomplish the difficult task
of deconvolving the image from imperfect and  noisy data. An objective
criterion  has therefore been established  to check the quality of the
results. A natural product of  the MCS code is  the RM (Residual Map),
which   is the  difference  between the   deconvolved  and real images
divided by  the    standard deviation of  each   pixel.   Whenever the
simultaneous deconvolution capabality of the method is used, as in the
present  case, the program returns  one RM for  each frame in the data
set.  A perfect deconvolved image should match the data at best in the
sense of the $\chi^{2}$ and leave flat RM with a mean  value of 1 (one
standard deviation) all over the field.

RMs guide the user in his/her choice of the smoothing to be applied to
the image in order to avoid local  under- or over-fitting of the data.
Still, on the  basis of the RMs, the  user can constrain the number of
point  sources  to be  involved   in the deconvolution.  For  example,
missing a bright point source  (i.e., with high S/N)  will result in a
``hot spot'' well above the  critical value of 1 in  the RMs.  This is
no longer true for very faint sources. In this case, no discrimination
can  be made between  an  extended object  and a point  source, so the
faint  point  sources   are  modelled as   part   of  the  deconvolved
background.  We therefore always choose   the minimum number of  point
sources  needed  in order to  produce  acceptable RMs. Thus, different
images  of the  same field can  have different   aspects, depending on
whether individual objects were modeled as point sources or not.

Objects  near the frame  edges   are  not well-fitted, especially   in
crowded  fields such as these  (e.g., objects labelled 1  to 4 in Fig.
1).  Object 1 leaves particularly significant structures in the RMs of
the $I$-band data, either when modeled as a point source or as part of
the  background in the deconvolved  image.  We therefore conclude that
it is extended or that it is a very strong blend of point sources.

In order to  run the MCS algorithm, the   user has to provide  initial
positions  and  intensities for  all points  sources.  The astrometric
1$\sigma$ error  bars were estimated as  the dispersion of the results
of several  deconvolutions using  different  initial  conditions.  The
typical astrometric  accuracy  is 0\farcs05\, for the  brightest point
sources, while it can be as much as 0\farcs2\,  for the faintest ones.
Since the peak intensity of a point source is  allowed to be different
in each data frame, the photometric error bar is simply the dispersion
of the peak intensities  found  after simultaneous deconvolution of  a
whole data set.  In order to quantify the additional errors introduced
by a poor knowledge of the PSF in the Keck $I$-band, the deconvolution
program was  run using 5 different  PSFs computed from different stars
in the field.

\section{Results and discussion}

In Fig.  1 and Plate 1 we present both  the raw and deconvolved images
in  $I$, $K^{\prime}$  and  $K$,  with a  resolution  of  the order of
0\farcs15-0\farcs20. A   red point source  is clearly  detected at the
position expected for the  NE radio source  of PKS~1830--211.  Another
red object is observed close to the position of the SW radio source of
the lensed system, but the extended nature of  the source and the poor
quality of  the PSF do not allow  to sort out its detailed morphology.
The photometry  and astrometry of the field  are presented in Tables 1
and 2 and in Fig.  2, along with our  estimates of the 1$\sigma$ error
bars.

 The red star  in D92 dominates the total  flux in the $I$-band, while
in the infrared  the component near  the NE  radio source takes  over.
With our  high S/N we can  show that  its shape  is compatible  with a
point source,  and  its color,   $I-K=6.9$, is  much   redder than any
``normal'' star (e.g.,  Koornneef, 1983).  The  $I$, $K^{\prime}$, and
$K$ positions are all within the 1$\sigma$ radio error bars. In Fig. 2
all  objects are aligned with  respect  to the M-star   so that we can
compare the different positions measured for the  NE and SW components
at optical and IR wavelengths.

 Molecular absorption   towards the two    lensed images was spatially
resolved at millimeter  wavelengths,  and the separation   between the
lensed images  was found to be 0\farcs98  (Frye et al.  1997).  In our
optical and IR images the SW  component is $0.61\pm 0.13$\arcsec \ and
$0.85\pm 0.09$\arcsec \  respectively, away from  the NE component.  A
plausible explanation for  the apparent  positional shift between  the
optical, IR and radio positions is that the SW component is a blend of
two objects: the lensing galaxy and  the heavily reddened SW component
seen in the radio  images.  This has been  very recently  confirmed by
HST NICMOS3  observations by the CASTLES  group (Kochanek et al. 1998)
which show the lensing  galaxy in  the center  of the ``hole''  of the
radio  ring, as predicted   by LensClean  modeling   of the system  by
Kochanek \& Narayan (1992). Note also  the very good general agreement
between  the HST  images and our  deconvolutions,  although the latter
were carried out without prior knowledge of the HST data.

 The flux ratio between the two  lensed images of  the source is 0.973
$\pm$    0.016 at  radio  wavelengths   (S90), and  about  18.2 in  $K$
(3.15$\pm$0.2 magnitudes).  The   combination of a  reddened  SW radio
source plus blue lens can explain the  large flux ratios.  Both NE and
SW components are  reddened, making PKS~1830--211 another good example
of a dusty lens, along with MG~0414+0534  (e.g., Annis \& Lupino 1993)
and  MG1131+0456 (Larkin et al.  1994),  the  mean galactic extinction
being far below  the values obtained for  PKS~1830--211.  At the  lens
redshift, the observed $K$-band corresponds to a central wavelength of
1.17  $\mu$ in the restframe.  For  the  galactic reddening curve, the
extinction   in this band is   $A_{(1.2\mu)}  = 0.37A_{V}$ (Sandage \&
Mathis, 1981).   Thus,  the   differential  extinction is  about   8.5
magnitudes  in the $V$-band and implies  $E(B-V)=2.75\pm 0.2$, in good
agreement with the  value of 2.4  independently  derived from  the HST
observations (Kochanek et al.  1998).

 If the total extinction is  comparable to the differential extinction
between the two images, then the source is  attenuated in the observed
$K$-band by about 3 magnitudes.  On the other hand, considering simple
SIS or  point-mass  models,   the magnification  by  the lens   may be
similar, or somewhat less, maybe 2 magnitudes.  So, a reasonable value
for the unobscured un-lensed $K$ magnitude of the source may be around
14.   A  typical   quasar  has $V-K\sim3$.    So   the unobscured $V$
magnitude  of the  source may be  about  17, which is  bright, but not
unreasonable for $z\sim 1-2$.

\section{Conclusion}

The main  result of our   study is the detection   of the optical  and
near-IR counterpart  to  the NE radio  source of  PKS~1830-211 and the
possible  detection of the SW component  and lensing galaxy.  However,
our SW component candidate might be the lensing galaxy alone or, given
the crowding in  the field,  a  red galactic object almost  coincident
with   the position of the  SW   radio source.  The   hypothesis of  a
demagnified third image of the source (S90)  between the 2 main lensed
images is unlikely as in such a case extinction of the lens would have
made it visible  in the IR.   Furthermore, the  IR centroid of  the SW
component  would have been shifted   towards ``object E'' rather  than
towards the radio position of the SW component.

The higher contrast between the NE  component and nearby M-star in the
infrared make near-IR  spectroscopy  necessary for finding  the source
redshift.  Deep, high resolution near-IR  imaging is needed to  reveal
the  exact nature of the  faint  SW component.   However, even at  the
highest resolution attainable,  0\farcs2- 0\farcs3 in  the IR with HST
(in particular in $K$  where the SW  component of PKS~1830-211 is best
visible), deconvolution will be  essential to discriminate between the
SW  component candidate,    the lensing galaxy  and   additional faint
galactic stars.


\acknowledgments

We  thank Jack Welch, Hy Spinrad,  Jens Hjorth, Andreas Jaunsen, Chris
Kochanek and the anonymous referee for useful discussions and comments
about the  first  version of   the manuscript.  We also   thank Alfred
Rosenberg who provided the ESO $I$-band frame used for the photometric
calibration.  The expert help of the staff at  ESO and WMKO during the
observing   runs  was  very  much  appreciated.  FC   is  supported by
contracts   ARC  94/99-178 ``Action  de   Recherche Concert\'ee  de la
Communaut\'e        Fran\c{c}aise''         and   P\^ole  d'Attraction
Interuniversitaire  P4/05  (SSTC, Belgium).   SGD and  MAP acknowledge
support from the NSF PYI award AST-9157412. Additional support for MAP
was provided by NASA through grant  number HF-01099.01-97A from STScI,
which is operated by AURA under NASA contract NAS5-26555.


%
%
\pagebreak

\pagebreak

\begin{figure}
\begin{center}
\leavevmode   
\epsfxsize=16cm   
\epsffile{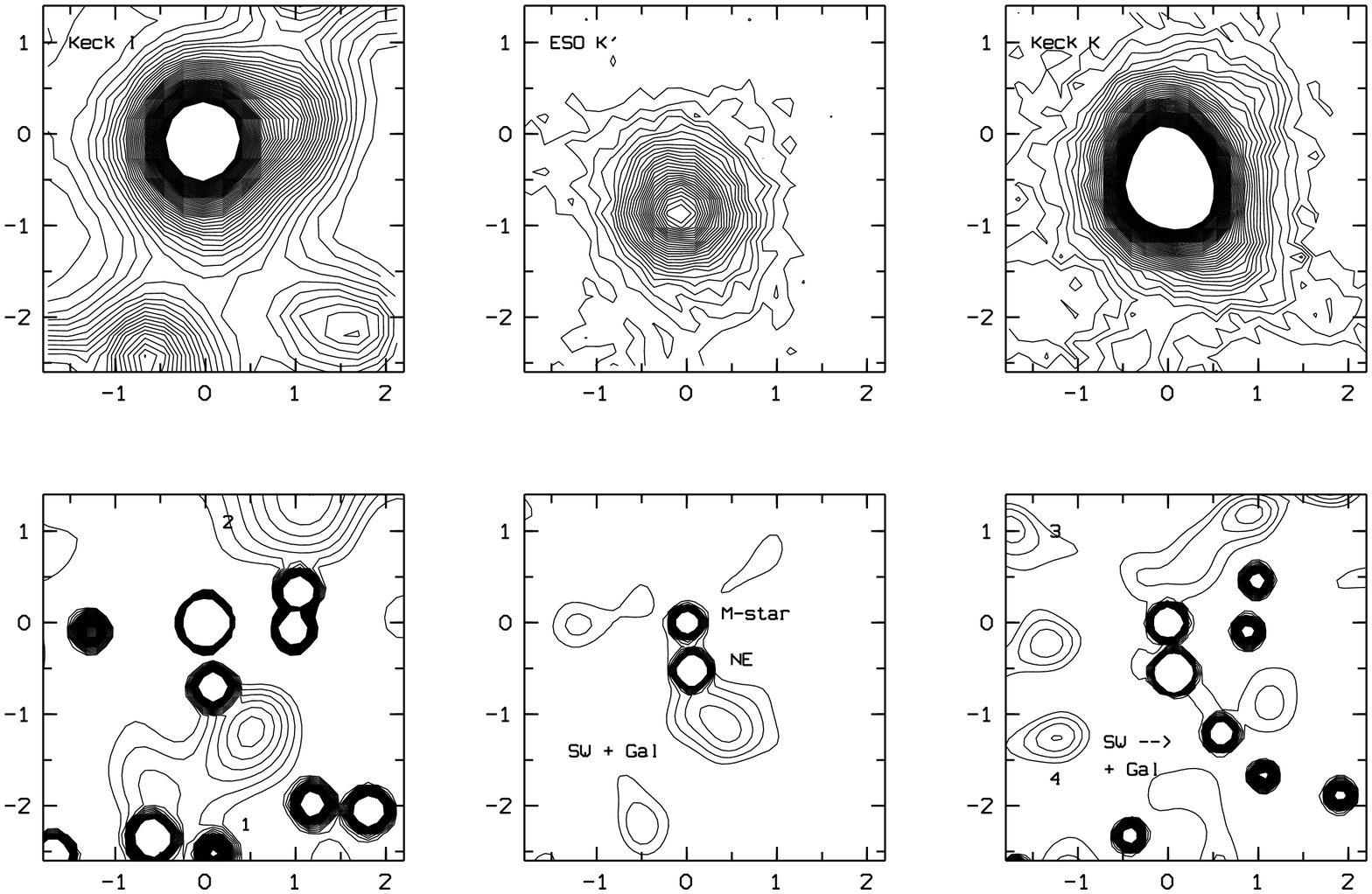}   

\figcaption{Four arcseconds field around PKS~1830--211 observed in the
$I$,  $K'$  and $K$ bands at  Keck  and ESO observatories.  [Top left]
Stack of 6 $I$-band frames with a pixel size  of 0\farcs215 and seeing
of 0\farcs8.   [Top middle] Mean  of 4   temporary stacks  (see  text)
obtained in  the $K^{\prime}$  band  with the  ESO/MPI 2.2m telescope.
The pixel size is 0\farcs151 and  the seeing is 0\farcs7.  [Top right]
Mean of 5  $K$-band images obtained with  Keck~{\sc i}  and NIRC.  The
pixel size  is 0\farcs157 and the seeing  is 0\farcs7.  [Bottom right]
Simultaneous deconvolution  of the 6   $I$-band frames: resolution  of
0\farcs215  and  pixel    size   of  0\farcs1075.    [Bottom   middle]
Simultaneous   deconvolution  of  the  4  $K^{\prime}$   images: final
resolution of 0\farcs151 and pixel size of 0\farcs075.  [Bottom right]
Deconvolution of the mean  of 5 NIRC  images: resolution of 0\farcs157
and pixel size of 0\farcs0785.  In all the images North is up, East to
the left.  The M-star, the NE QSO component  candidate and the blended
SW  component + lensing galaxy candidate  are indicated. In all images
the first countour corresponds to  2.5$\sigma_{sky}$. The step between
two contours is 2.5$\sigma_{sky}$.}

\label{fig:deconv}
\end{center}
\end{figure}

\begin{figure}
\begin{center}
\leavevmode   
\epsfxsize=10cm   
\epsffile{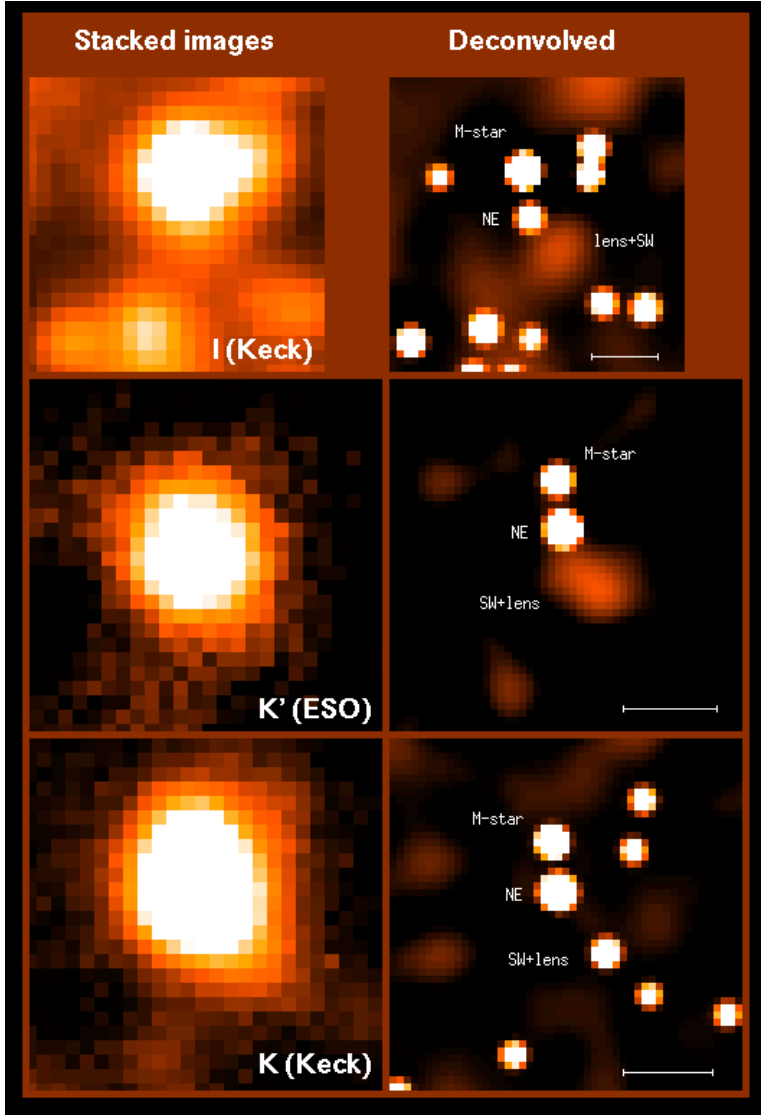}   

\figcaption{Field  around PKS~1830--211 observed  in the $I$, $K'$ and
$K$  bands  at Keck  and  ESO observatories.  [Top  left]   Stack of 6
$I$-band  frames  with a   pixel  size of   0\farcs215 and  seeing  of
0\farcs8.    [Middle left] Mean  of    4 temporary stacks (see   text)
obtained in  the  $K^{\prime}$ band with  the  ESO/MPI 2.2m telescope.
The pixel  size is  0\farcs151  and the seeing   is 0\farcs7.  [Bottom
left] Mean of  5 $K$-band images obtained with  Keck~{\sc i} and NIRC.
The pixel size is 0\farcs157 and the seeing  is 0\farcs7.  [Top right]
Simultaneous deconvolution   of the 6  $I$-band  frames: resolution of
0\farcs215 and pixel size of 0\farcs1075.  [Middle right] Simultaneous
deconvolution   of  the 4 $K^{\prime}$  images:    final resolution of
0\farcs151 and pixel size of 0\farcs075.  [Bottom right] Deconvolution
of the mean of 5 NIRC images: resolution  of 0\farcs157 and pixel size
of 0\farcs0785.  In all the images North is up, East to the left.  The
M-star, the NE QSO component candidate and  the blended SW component +
lensing galaxy  candidate are indicated.   The horizontal line in each
deconvolved image is   1\arcsec\, long.  The  cuts of  the  images are
chosen in order to display the full dynamics of the image at low light
levels.}

\label{fig:deconv}
\end{center}
\end{figure}

\begin{figure}
\begin{center}
\leavevmode 
\epsfxsize=14cm  
\epsffile{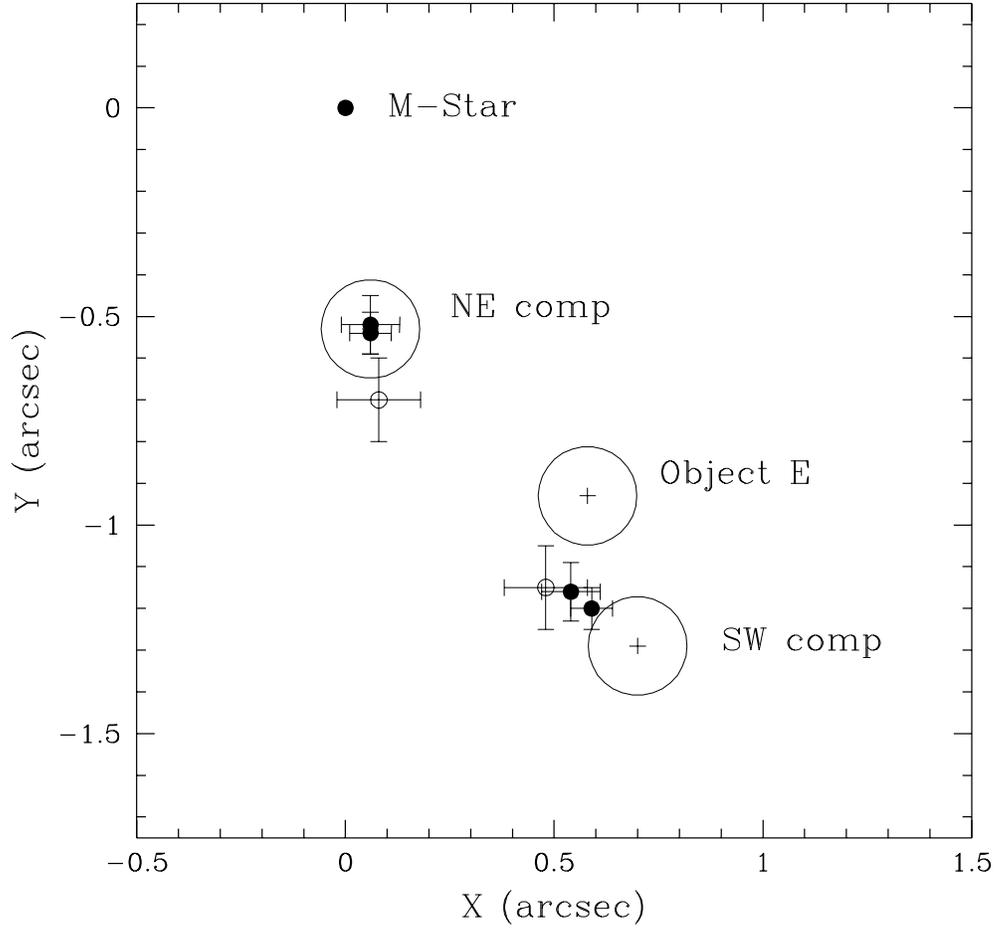}  
\figcaption{Positions
observed for the different objects detected in the optical and near-IR
along with their 1$\sigma$  error bars, relative to the  M star.   The
large open circles are radio positions whose radius corresponds to the
error bars quoted by S90.  The  black dots show the positions measured
from the  near-IR  images   while the  open circles  show   the result
obtained from the $I$-band data.}
\label{fig:geom}
\end{center}
\end{figure}

\begin{table}[t]
\begin{center}
\caption{Summary of the photometry.}
\vspace*{10mm}
\begin{tabular}{c c c c c}
\hline \hline 
& M-star & NE Comp. & SW Comp.+Lens & Limiting Mag.\\ \hline
$I(Keck)$ & $19.3 \pm 0.1$ & $22.0 \pm 0.2$ & $22.3 \pm 0.3$ & 24.0 \\
$J(ESO)$  & $18.1 \pm 0.1$ & $18.7 \pm 0.3$ & $>$ 20.5 & 20.5 \\ 
$K^{\prime}(ESO)$ & $17.3 \pm 0.3$ & $15.8 \pm 0.2$ & $19.0 \pm 0.4$& 19.1 \\ 
$K(Keck)$ & $16.6 \pm 0.2$ & $15.1 \pm 0.1$ & $18.2 \pm 0.2$ & 21.3 \\ 
\hline
\end{tabular}
\tablecomments{Results are given for the M-Star, the NE component, and
the SW   component  of the  lensed  source plus  lensing  galaxy.  The
limiting magnitude (point sources) for each  of the bands is given
in the final column.}
\end{center}
\end{table}

\begin{table}[t]
\begin{center}
\caption{Summary of the astrometry.}

\vspace*{10mm}
\begin{tabular}{c c c}
\hline \hline 
& NE Comp. & SW Comp.+Lens \\ \hline 
x($I$)& $+0.08 \pm 0.1$ & $ +0.48 \pm 0.1$\\ 
y($I$)& $-0.70 \pm 0.1$ & $ -1.15 \pm 0.1$\\ 
\hline 
x($J$)& $-0.07 \pm 0.1$ & $-$ \\ 
y($J$)& $+0.54 \pm 0.1$ & $-$ \\ 
\hline 
x($K^{\prime}$)& $ +0.06 \pm 0.07$ & $+0.54 \pm 0.07$\\
y($K^{\prime}$)& $ -0.52 \pm 0.07$ & $-1.16 \pm 0.07$\\ 
\hline
x($K$)& $+0.06 \pm 0.05$ & $+0.59 \pm 0.05$\\ 
y($K$)& $-0.54 \pm 0.05$ & $-1.20 \pm 0.05$\\ 
\hline
\end{tabular}
\tablecomments{Results  for  the NE and SW  components   of the lensed
source  plus lensing galaxy are  given in  arcseconds, relative to the
M-star, together with their 1$\sigma$ error bars, }
\end{center}
\end{table}


\begin{references}

\reference{} Annis, J.E., \& Luppino, G. 1993, ApJ, 407, L69

\reference{} Bessell, M. S., \& Brett, J. M. 1988, PASP, 100, 1134

\reference{} Combes, F., \& Wiklind, T. 1997, preprint astro-ph/9711184

\reference{} Courbin, F., Lidman, C., \& Magain, P.  1998, A\&A, 330, 57

\reference{} Djorgovski, S. G., Meylan, G., Klemola, A. et al. 
1992, MNRAS, 257, 240

\reference{} Frye, B. L., Welch, W. J. W., \& Broadhurst, T. J.
 1997, ApJ, 478, L25

\reference{} Jauncey, D. L., Reynolds,   J. E., Tzioumis, A. K. et  al. 1991,
Nature, 352, 132

\reference{} Kochanek, C.   S., Falco, E.E.,  Impey,  C. et al.  1998,
``The CASTLES Web pages'': http://cfa-www.harvard.edu/castles/1830.html

\reference{} Kochanek, C.S., Narayan, R., 1992, ApJ, 401, 461

\reference{} Koornneef, J. 1983, A\&A, 128, 84

\reference{} Larkin, J. E., Matthews, \& K., Lawrence, C. R. 1994, ApJ, 420, L9

\reference{} Magain, P.,  Courbin, F., \& Sohy, S. 1998,  ApJ, 494, 472

\reference{} Matthews, K., \&  Soifer, B. T. 1994,  in: ``Infrared Astronomy
with Arrays, ed. I. McLean, Dordrecht: Kluwer, p 239

\reference{} Oke, J. B. et al. 1995, PASP, 107, 375

\reference{} Savage, B., \& Mathis, J. S. 1981, ARA\&A, 17, 73

\reference{}  Subrahmanyan, R.,  Narasimha, D.,  Rao,  P., Swarup,  G. 1990,
MNRAS, 246, 263

\reference{}   van Ommen,  T. D.,  Jones,    D. L., Preston, R. A., Jauncey,
D. L. 1995, ApJ, 444, 561

\reference{} Wiklind, T., \& Combes, F. 1996, Nature, 379, 139

\end{references}
\end{document}